\documentclass[reprint, amsmath, amssymb,  superscriptaddress, aps, prd]{revtex4-2}
\usepackage{graphicx}
\usepackage{dcolumn}
\usepackage{bm}
\usepackage{hyperref}
\hypersetup{
colorlinks=true, 
colorlinks = true,
linkcolor = blue,
urlcolor  = blue,
citecolor = blue,
anchorcolor = blue}
\usepackage[mathlines]{lineno}

\begin{document}

\title{Observation of the cosmic ray shadow of the Sun with the ANTARES neutrino telescope}

\author{A.~Albert}
\affiliation{\scriptsize{Universit\'e de Strasbourg, CNRS,  IPHC UMR 7178, F-67000 Strasbourg, France}}
\affiliation{\scriptsize Universit\'e de Haute Alsace, F-68200 Mulhouse, France}

\author{M.~Andr\'e}
\affiliation{\scriptsize{Technical University of Catalonia, Laboratory of Applied Bioacoustics, Rambla Exposici\'o, 08800 Vilanova i la Geltr\'u, Barcelona, Spain}}

\author{M.~Anghinolfi}
\affiliation{\scriptsize{INFN - Sezione di Genova, Via Dodecaneso 33, 16146 Genova, Italy}}

\author{G.~Anton}
\affiliation{\scriptsize{Friedrich-Alexander-Universit\"at Erlangen-N\"urnberg, Erlangen Centre for Astroparticle Physics, Erwin-Rommel-Str. 1, 91058 Erlangen, Germany}}

\author{M.~Ardid}
\affiliation{\scriptsize{Institut d'Investigaci\'o per a la Gesti\'o Integrada de les Zones Costaneres (IGIC) - Universitat Polit\`ecnica de Val\`encia. C/  Paranimf 1, 46730 Gandia, Spain}}

\author{J.-J.~Aubert}
\affiliation{\scriptsize{Aix Marseille Univ, CNRS/IN2P3, CPPM, Marseille, France}}

\author{J.~Aublin}
\affiliation{\scriptsize{Universit\'e de Paris, CNRS, Astroparticule et Cosmologie, F-75006 Paris, France}}

\author{B.~Baret}
\affiliation{\scriptsize{Universit\'e de Paris, CNRS, Astroparticule et Cosmologie, F-75006 Paris, France}}

\author{S.~Basa}
\affiliation{\scriptsize{Aix Marseille Univ, CNRS, CNES, LAM, Marseille, France }}

\author{B.~Belhorma}
\affiliation{\scriptsize{National Center for Energy Sciences and Nuclear Techniques, B.P.1382, R. P.10001 Rabat, Morocco}}

% ...

\collaboration{ANTARES Collaboration}

\date{\today}

\begin{abstract}

The ANTARES detector is an undersea neutrino telescope in the Mediterranean Sea. The search for point-like neutrino sources is one of the main goals of the ANTARES telescope, requiring a reliable method to evaluate the detector angular resolution and pointing accuracy. This work describes the study of the Sun ``shadow" effect with the ANTARES detector. The shadow is the deficit in the atmospheric muon flux in the direction of the Sun caused by the absorption of the primary cosmic rays. This analysis is based on the data collected between 2008 and 2017 by the ANTARES telescope. The observed statistical significance of the Sun shadow detection is $3.7\sigma$, with an estimated angular resolution of $0.59^\circ\pm0.10^\circ$ for downward-going muons. The pointing accuracy is found to be consistent with the expectations and no evidence of systematic pointing shifts is observed.
\end{abstract}

\maketitle

\section{Introduction}


Charged cosmic rays (mainly protons), $\gamma$-rays and neutrinos represent relevant probes for high-energy astrophysical research. However, $\gamma$-rays with energies higher than few TeV interact with the infrared and the cosmic microwave background producing electron-positron pairs. Charged cosmic rays (CRs) are deflected by cosmic magnetic fields and it is almost impossible to identify their origin through the measurement of their arrival direction. Moreover, the structure of galactic magnetic fields is so complex that the distribution of galactic CRs is almost isotropic near the Earth. Neutrinos have properties which allow to observe and study the Universe in a unique way. They can propagate from their sources to the Earth without changing trajectory and with small probability of being absorbed. 

The ANTARES undersea neutrino telescope \citep{Main} is primarily designed for the detection of neutrino point-like sources and both the pointing accuracy and the angular resolution of the detector are important for the evaluation of the telescope performance.

The interaction of primary CRs in the atmosphere produces secondary downward-going muons that can be detected in the undersea detector. However, the CRs could be absorbed by the Moon and the Sun leading to a deficit in the atmospheric muon flux in the directions of these celestial bodies. This effect has been observed by several experiments: CYGNUS \cite{cyg}, TIBET \cite{tib}, BUST \cite{bust}, CASA \cite{casa}, MACRO \cite{macro}, SOUDAN \cite{soudan}, ARGO-YBG \cite{argo}, HAWC \cite{HAWC}, MINOS \cite{minos} and also IceCube \cite{ice}. A Moon shadow analysis with the ANTARES telescope, corresponding to a total livetime of 3128 days, has also been published \cite{Moon_Antares}.

This work presents the Sun shadow analysis using the ANTARES 2008-2017 data sample, corresponding to a total detector livetime of 2925 days. The analysis is based on $2.6\times10^6$ events reconstructed as downward-going muons with the standard ANTARES reconstruction chain \cite{reconstructed}.

The paper is organised as follows: in Section 2 the ANTARES neutrino telescope is described; the Sun shadow analysis and the obtained results are presented in Section 3; finally, the conclusions are summarised in Section 4.

\section{The ANTARES neutrino telescope}

The ANTARES undersea neutrino telescope is taking data in its final configuration since 2008. It is located in the Mediterranean Sea, 40 km offshore from Toulon (France) at 42$^\circ$48’ N latitude and 6$^\circ$10’ E longitude. The detector consists of twelve lines, each is about 450 m long. Each line comprises 25 storeys with three 10-inch photomultiplier tubes (PMTs) inside pressure resistant glass spheres (the optical modules). The first instrumented storey is located 100 m above the seabed. The distance between storeys is 14.5 m and the distance between two lines is about 65 m. The lines are connected to a junction box that links the detector to the shore station through an electro-optical cable about 40 km long. 

A relativistic muon induces Cherenkov photons when travelling through the water, which are detected by the PMTs producing a signal (\textit{hit}) \cite{Main}. The PMTs face $45^{\circ}$ downward in order to optimise the detection of light from upward-going particles. The set of hits detected within a certain time window is called \textit{event}. If the hits of one event satisfy spacetime causality the event is identified as a muon candidate \cite{reconstructed, muon_cand}. The reconstruction of the tracks is based on the probability density function of the arrival times of photons at the PMTs.

\section{The Sun shadow analysis}

The ANTARES telescope can detect only downward-going atmospheric muons because the upward-going ones are absorbed by the Earth. The energy threshold of muons at the sea surface level that can reach the detector is about 500 GeV \cite{energy_threshold}. In this energy range the direction of primary CRs may be assumed as collinear with the secondary muons. Even though the solar magnetic field is not expected to introduce a systematic shift in the pointing accuracy derived using the Sun shadow effect, it is expected that its influence can lead to a blurring of the shadow \cite{ice}. Therefore, the primaries that are blocked by the Sun lead to a deficit in the atmospheric muon flux in the direction of the Sun.
 
The analysis is performed in three steps. The first one is the data selection optimisation which provides the best sensitivity for the observation of the deficit of events from the direction of the Sun. The second step provides the estimation of the angular resolution of the detector for the reconstructed downward-going events. And in the third step a possible shift of the Sun shadow centre with respect to the nominal Sun position is investigated using a two-dimensional approach. 

\subsection{The data selection optimisation} 
\label{MC significance}

A Monte Carlo (MC) simulation is produced and exploited in order to optimise the event selection criteria of the analysis. The simulation features downward-going muon events which are generated at the detector level with the MUPAGE code \cite{mupage}. MUPAGE is based on parametric formulas that allow to calculate the flux and the angular distribution of underwater muon
bundles, taking into account the muon multiplicity and the energy spectrum. Muons are generated on the surface of a cylinder (\textit{can}) surrounding the active volume of the detector, 650 m high, with a radius of 290 m. The simulation includes the propagation of the muons in the instrumented volume, the induced emission of Cherenkov light, the light propagation to the optical modules and the digitised response of the PMTs \cite{mupage_description}. In order to reproduce the time variability of the detector conditions, the MC sample is subdivided in batches corresponding to the actual data-taking periods (run-by-run MC simulation \cite{RbR}). The trade-off between the accuracy of the simulations and CPU time, exploited to produce the MC sample, limited the MC muon statistics to 1/3 of the actual expected one. In order to enlarge the statistics of MC simulation the \textit{additional zones} approach was performed: since the muon generation is produced on a full-sky base, regions of the sky with the same occupancy as of the Sun region can be exploited, where the MC sample can be increased. Additional zones are obtained artificially by shifting the Sun position by 2, 4, 6, 8, 12, 14, 16, 18, 20 and 22 hours. Therefore, the whole statistics of MC considering 11 additional zones together with the Sun zone is 4 times larger than the real data statistics. 

Since the atmospheric muon flux is not uniform, the distribution of muons depends also on the Sun elevation angle at the moment of muon detection. Statistics of the events significantly decreases for muon tracks close to the horizon, for this reason a cut on the Sun elevation angle is applied: $\theta_{\rm{Sun}}>15^\circ$. 

The quality of the reconstructed tracks is determined by two parameters: the likelihood-wise parameter, $\Lambda$, and the angular error estimator of the reconstructed direction, $\beta$ \cite{quality}. In order to determine for which set of cut values on $\Lambda$ and $\beta$ the sensitivity of the Sun shadow detection is maximal, the hypothesis test approach is used. The null hypothesis $H_0$ corresponds to the absence of the Sun shadowing effect, while the $H_1$ hypothesis is compliant with the presence of this phenomenon. Then two different MC samples are generated. According to the null hypothesis, in the first sample the Sun shadow effect is not introduced in the simulation; according to the $H_1$ hypothesis, in the second sample the Sun shadow effect is obtained by removing all the muons generated within the Sun disk, assuming a radius of $0.26^\circ$. For each sample, the distribution of events as a function of the angular distance from the Sun, up to 10$^\circ$, is produced. Such a histogram is subdivided into 25 bins with size $\Delta\delta=0.4^\circ$. Each bin corresponds to a concentric ring with increasing radius centred on the Sun position. The content of each bin is normalised to the corresponding area of the ring, resulting in an event density. 

Assuming that the event population in each bin asymptotically follows a Gaussian probability distribution, the test statistic is calculated under the above mentioned two hypotheses as a $\chi^2$ difference, resulting in $\lambda_{0}$ and $\lambda_{1}$:

\begin{equation}
\begin{aligned}
\lambda_{0}=\sum_{i=1}^{N_{\rm{bins}}} [\frac{(n_0^i - \mu_i)^2}{\sigma^2_{\mu,i}} - \frac{(n_0^i - \nu_i)^2}{\sigma^2_{\nu,i}}],
\\
\lambda_{1}=\sum_{i=1}^{N_{\rm{bins}}} [\frac{(n_1^i - \mu_i)^2}{\sigma^2_{\mu,i}} - \frac{(n_1^i - \nu_i)^2}{\sigma^2_{\nu,i}}],
\end{aligned}
\end{equation}

\noindent with $\mu_i$ ($\nu_i$) the expected number of events in the i-th bin under $H_1$ ($H_0$) hypothesis, $\sigma_{\mu,i}$ ($\sigma_{\nu,i}$) the error in the i-th bin under $H_1$ ($H_0$). The values of $n_1$ ($n_0$) are derived according to a Poisson distribution with expectation values equal to $\mu_i$ ($\nu_i$). A total of $10^6$ pseudo-experiments are generated to build the distribution of the test statistic.
 
\begin{figure}[htbp]
\begin{center}
\centering
\includegraphics[width=0.5\textwidth]{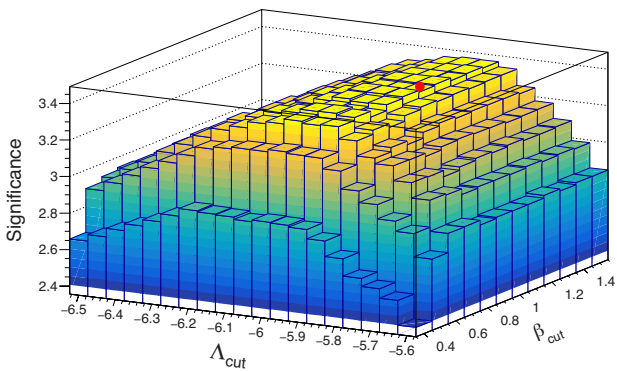}
\caption{Expected statistical significance of the Sun shadow detection during the period from 2008 to 2017 based on MC simulations, as a function of cut values on $\Lambda$ and $\beta$ ($\Lambda_{\rm{cut}}$ and $\beta_{\rm{cut}}$). The red point represents the selected set of cut values ($\Lambda_{\rm{cut}}=-5.9$ and $\beta_{\rm{cut}}=1.1^{\circ}$). The expected significance for the selected set of cut values is $3.4\sigma$.}
\label{fig:significance_and_cuts}
\end{center}
\end{figure}

\begin{figure}[htbp]
\begin{center}
\centering
\includegraphics[width=0.5\textwidth]{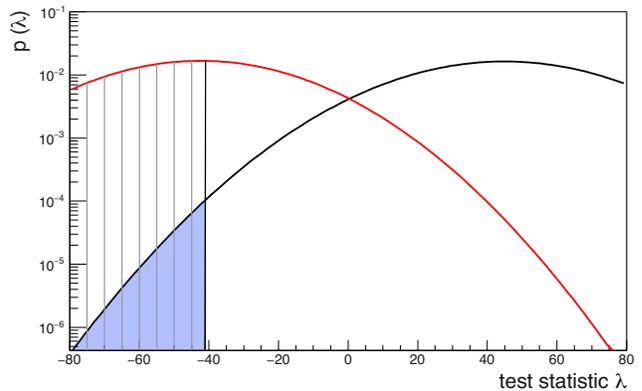}
\caption{Distribution of the test statistic $\lambda$ for the two hypotheses, $H_0$ (black curve) and $H_1$ (red curve), obtained for the optimized set of cut values ($\Lambda_{\rm{cut}}=-5.9$ and $\beta_{\rm{cut}}=1.1^{\circ}$). The dashed area represents the fraction of pseudo-experiments ($50\%$) where $H_1$ hypothesis is correctly identified. The coloured area corresponds to the expected median significance ($3.4\sigma$) to reject the $H_0$ hypothesis in favour of the $H_1$ hypothesis.}
\label{fig:lambda_curves}
\end{center}
\end{figure}

The hypothesis test procedure is repeated for different sets of cut values on $\Lambda$ and $\beta$ to maximise the sensitivity to the Sun shadow detection (Fig. \ref{fig:significance_and_cuts}). Fig. \ref{fig:lambda_curves} shows the estimation of the sensitivity. It is evaluated through the computation of the p-value of the $\lambda_0$ distribution (null hypothesis, $H_0$) corresponding to the median of the $\lambda_1$ distribution, for which $50\%$ of the pseudo-experiments under the $H_1$ hypothesis (presence of the Sun shadow) are correctly identified. For the optimised values of $\Lambda$ and $\beta$, the p-value is equal to $7.4 \times 10^{-4}$, corresponding to a significance of $3.4\sigma$.

It is found that the sensitivity is almost constant for $-6.0<\Lambda_{\rm{cut}}<-5.9$ and $0.6^\circ<\beta_{\rm{cut}}<1.5^\circ$. In this parameter space, a particular set of cut values is chosen: $\Lambda_{\rm{cut}}=-5.9$ and $\beta_{\rm{cut}}=1.1^{\circ}$. For this set of cut values the muon density far from the Sun position is flat, this condition is required in the data significance estimation approach that will be described below in Sect. \ref{Angular resolution}. 

\subsection{The angular resolution estimation and significance of the results}
\label{Angular resolution}

The reconstructed events from the 2008-2017 ANTARES data sample are selected with the optimised cut values described above, providing $6.5\times10^5$ events. The data event density distribution is produced in the same way as for the MC events described above in the hypothesis test procedure. 

\begin{figure}[htbp]
\begin{center}
\centering
\includegraphics[width=0.5\textwidth]{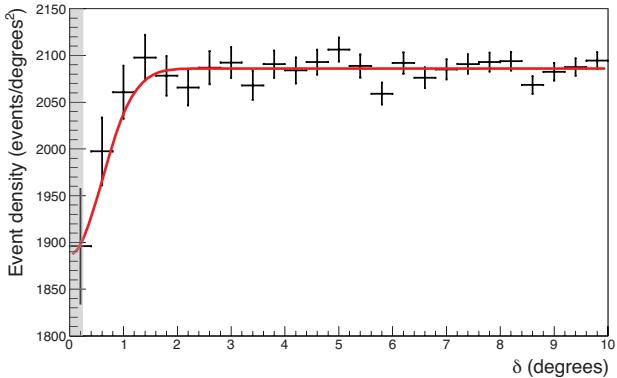}
\caption{The muon event density as a function of the angular distance $\delta$ from the Sun centre based on the data sample taken in period 2008-2017 fitted with Eq. \ref{eq:gaussian1} (red line). The shaded area corresponds to the Sun angular radius ($0.26^\circ$).}
\label{fig:data_fit}
\end{center}
\end{figure}

In order to estimate the angular resolution of the detector for downward-going muons, the data histogram is fitted with the following function \cite{Moon_Antares} (red line in Fig. \ref{fig:data_fit})

\begin{equation}
f(\delta)=\frac{dN}{d\Omega}=k(1-\frac{R^2_{\rm{Sun}}}{2\sigma_{\rm{res}}^2}e^{-\frac{\delta^2}{2\sigma_{\rm{res}}^2}}), 
\label{eq:gaussian1}
\end{equation}

\noindent where $\Omega$ is the solid angle of the concentric ring around the Sun centre, $k$ is the average muon event density in the $H_0$ hypothesis, the value of $k$ from the fit is $2086 \pm 2.6$, $R_{\rm{Sun}}$ is the average angular radius of the Sun ($0.26^\circ$) and $\sigma_{\rm{res}}$ is the width of the Gaussian dip. The number of absorbed events in the Sun shadow dip is $N_{\rm{abs}}=k \pi R^2_{\rm{Sun}}=443 \pm 0.6$. The average muon event density obtained in the ANTARES Moon shadow study is $2376 \pm 3$ \cite{Moon_Antares} and hence the number of absorbed events is $505 \pm 0.6$. The average muon event density in the Sun analysis is smaller with respect to the Moon analysis since the quality category to include a run has been slightly changed and a sample of runs with a tighter quality selection is chosen for the current analysis. 

The value of $\sigma_{\rm{res}}$ from the fit is $0.59^\circ\pm0.10^\circ$. The goodness of the fit is found to be $\chi^2/\rm{dof}=19.6/23$. 

Pseudo-experiments are used to evaluate the actual effect of a finite-size radius of the Sun. Several event densities are produced and convoluted with a step function representing the Sun radius assuming different detector angular resolutions. The discrepancies obtained between the assumed detector angular resolutions and the fitted values of the Gaussian width are below $10\%$ for the assumed angular resolution values above $0.35^{\circ}$, i.e. negligible with respect to the statistical uncertainty. Therefore, the obtained value of $\sigma_{\rm{res}}$ can be treated as the angular resolution of the telescope for downward-going muons. 

The angular resolution value $\sigma_{\rm{res}}$ is compatible with the one obtained in the ANTARES Moon shadow analysis ($0.73^\circ\pm0.14^\circ$), with a $3.5\sigma$ significance of lunar detection \cite{Moon_Antares}.

The statistical significance of the result is estimated using the hypothesis test approach. For the $H_0$ hypothesis no shadowing effect is assumed. Under this hypothesis the data event density in Fig. \ref{fig:data_fit} is fitted with the function which has one free parameter, $k$:

\begin{equation}
 \frac{dN}{d\Omega}=k;
\label{eq:flat}
\end{equation}

\noindent the corresponding $\chi^2$ value is $\chi^2_0=33.5$. The $H_1$ hypothesis corresponds to the presence of the shadowing effect according to Eq. \ref{eq:gaussian1}. The corresponding deviation of the data with respect to the $H_0$ hypothesis is computed by means of the test statistics: $-\lambda = \chi^2_0 - \chi^2_1$, which follows a $\chi^2$ distribution with 1 degree of freedom. A significance of $3.7\sigma$ is found. According to the MC pseudo-experiments the probability to obtain such value of significance or higher is $37\%$.

As reported by the IceCube Collaboration \cite{ice}, the primary CRs can be influenced by the Sun magnetic field which can lead to the blurring of the shadow. In order to study the influence of this effect, the data sample is divided into two samples with roughly equal statistics. The first one covers the period from the middle of 2008 to the middle of 2011, when the Sun activity was in the lower half, while the second one covers the period from the middle of 2011 to the end of 2015, when the Sun activity was in the higher half. The statistical significance of the Sun shadow observation is almost the same in both data samples, $2.6\sigma$ and $2.5\sigma$ for the first and the second data samples respectively, the spreading of the dip is also compatible within the statistical uncertainties. This is compatible with the results obtained in the other experiments since the statistics of the data sample is insufficient to obtain significative conclusions.

\subsection{Absolute pointing}

The procedure for the estimation of the pointing accuracy of the Sun shadow detection follows that used for the ANTARES Moon shadow study \cite{Moon_Antares}. The distribution of events which satisfies the selection criteria described previously is projected in a two-dimensional histogram as a function of $x=(\alpha_{\mu} - \alpha_{\rm{Sun}})\times \cos(h_{\mu})$ and $y=h_{\mu} - h_{\rm{Sun}}$, where $\alpha_{\mu}$, $\alpha_{\rm{Sun}}$ are the azimuthal coordinates and $h_{\mu}$, $h_{\rm{Sun}}$ are the elevation angles of the reconstructed track and the Sun, respectively. The histogram range is $[-10^\circ,10^\circ]$ for both $x$ and $y$, and it is divided in a grid of $0.4^\circ \times 0.4^\circ$ squared bins. 

For the determination of a possible shift of the Sun shadow centre with respect to the nominal Sun position the following approach is used. Since the atmospheric muon flux depends mainly on the elevation angle, in the $H_0$ hypothesis (no shadowing effect), the background distribution is approximated with a second degree polynomial: 

\begin{equation}
\label{SDP}
p_2(x,y;\textbf{k})=k_0 + k_1x + k_2x^2 + k_3y + k_4y^2.
\end{equation}

In the $H_1$ hypothesis (presence of the shadowing effect), the data distribution is approximated with a function obtained by subtracting from $p_2(x,y;\textbf{k})$ a two-dimensional Gaussian function:

\begin{equation}
G(x,y;A_{sh},x_s,y_s)=\frac{A_{sh}}{2\pi\sigma_{\rm{res}}^2}e^{-\frac{(x-x_s)^2 + (y-y_s)^2}{2\sigma_{\rm{res}}^2}},
\end{equation}

\noindent where $A_{sh}$ is the amplitude of the deficit caused by the Sun shadow (free parameter), $(x_s,y_s)$ is the assumed position of the Sun. The width of the Gaussian function is assumed to be the same in both dimensions, so that $\sigma_x=\sigma_y\equiv\sigma_{\rm{res}}$, and  $\sigma_{\rm{res}}$ is fixed to the value of the angular resolution defined in Eq. \ref{eq:gaussian1} and derived in the previous sub-section. 

In the pointing accuracy estimation, the Sun shadow centre is assumed to be in the different points of the two-dimensional histogram described above with a step size of $0.1^\circ$. The nominal Sun position is $O\equiv(0^\circ,0^\circ)$. The test statistic function is then calculated for each assumed shift of the Sun position as:

\begin{equation}
\lambda(x_s,y_s)=\chi^2_{H_1}(x_s,y_s) - \chi^2_{H_0},
\end{equation}

\noindent where $\chi^2_{H_0}$ is the $\chi^2$ value obtained from the fit with Eq. \ref{SDP}, which is a constant value for all the bins of the histogram, and $\chi^2_{H_1}(x_s,y_s)$ is the $\chi^2$ value obtained from the fit with the function used to describe hypothesis $H_1$, $p_2(x,y;\textbf{k}) - G(x,y;A_{sh},x_s,y_s)$.

\begin{figure}[htbp]
\begin{center}
\centering
\includegraphics[width=0.5\textwidth]{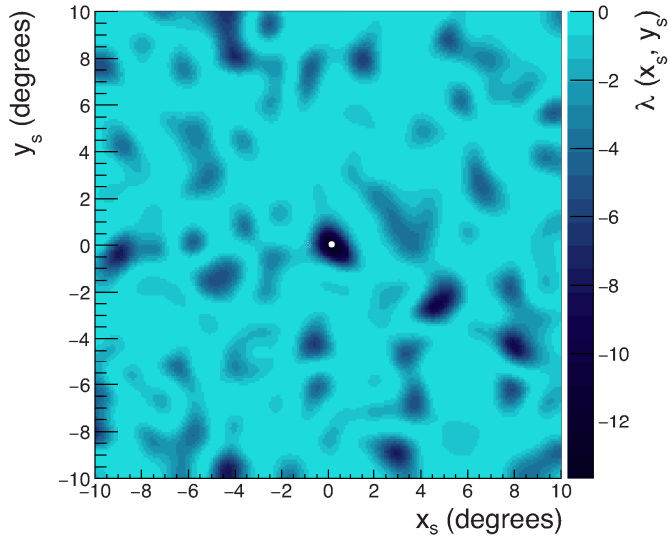}
\caption{The distribution of the test statistic values around the nominal Sun position $O\equiv(0^\circ,0^\circ)$. The minimum value $\lambda_{\rm{min}}=-13.7$ is found at $(0.2^\circ,0^\circ)$ point (white dot).}
\label{fig:delta-chi2}
\end{center}
\end{figure}

Fig. \ref{fig:delta-chi2} shows the values of the test statistic as a function of the assumed Sun position, $\lambda(x_s,y_s)$. The minimum value of $\lambda(x_s,y_s)$ is found at $(0.2^\circ,0^\circ)$ point and it is equal to $\lambda_{\rm{min}}=-13.7$. The corresponding fitted value of the Sun shadow dip amplitude is $A_{min}=55 \pm 15$. The values of $\lambda(x_s,y_s)$ and $A_{sh}$ for the nominal Sun position are $\lambda_{O}=-13.1$ and $A_{O}=54 \pm 15$. At each bin, $-\lambda$ follows the distribution of a $\chi^2$ with one degree of freedom, assuming $H_0$ as the true hypothesis. This allows the significance to reject the no-Sun hypothesis to be estimated. Considering $-\lambda_{O}$, a p-value of $3.1\times10^{-4}$ is obtained. The corresponding significance is $3.6\sigma$.

\begin{figure}[htbp]
\begin{center}
\centering
\includegraphics[width=0.5\textwidth]{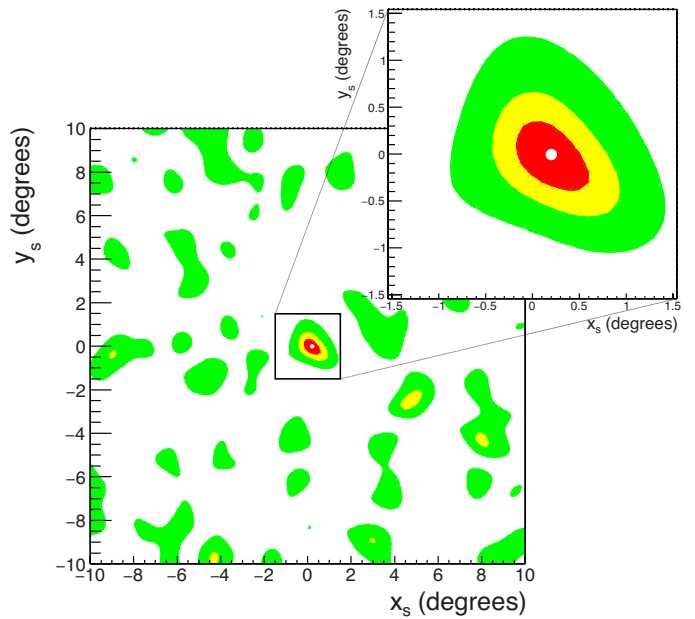}
\caption{Contours corresponding to different confidence levels (red: $68.27\%$; yellow: $95.45\%$; green: $99.73\%$). The white dot indicates $(0.2^\circ,0^\circ)$ point for which a minimum value of $\lambda_{\rm{min}}=-13.7$ is obtained.}
\label{fig:contours}
\end{center}
\end{figure}

The distribution of values of the test statistic $\lambda(x_s,y_s)$ can be interpreted as a bi-dimensional profile-likelihood, with $A_{sh}$ treated as the nuisance parameter. Therefore, the interval corresponding to a desired confidence level (CL) is obtained for $\lambda(x_s,y_s) \leq \lambda_{\rm{cut}} = \lambda_{\rm{min}} + Q$, where $Q$ is the quantile for the joint estimation of two parameters, according to the values reported on Table 40.2 of \cite{PDG}. Fig. \ref{fig:contours} shows the estimation of the confidence regions for CL $\equiv\lbrace68.27\%, 95.45\%, 99.73\%\rbrace$.

\section{Conclusions}

The evaluation of the angular resolution of the ANTARES detector is essential since one of the main goals of the telescope is the search for point-like sources \cite{point-source1, quality, point-source2}.

This paper presents the observation of the Sun shadow with the ANTARES neutrino telescope. The analysis is based on the data taken in the period between 2008 and 2017 corresponding to a total detector livetime of 2925 days. 

The Sun shadow effect is studied by means of two complementary approaches which allow to determine the angular resolution for downward-going atmospheric muons and to verify the pointing performance of the detector. The shadow effect is observed with $3.7\sigma$ statistical significance using the one-dimensional approach. The angular resolution for downward-going muons is found to be equal to $0.59^\circ\pm0.10^\circ$. A better angular resolution is expected for upward-going events, as the PMTs of the detector are pointing $45^{\circ}$ below the horizon to maximize the light collection for upward-going neutrino-induced events. 

The obtained angular resolution is compatible with the angular resolution found in the Moon shadow analysis with the ANTARES telescope ($0.73^\circ\pm0.14^\circ$) \cite{Moon_Antares}.

The influence of the Sun magnetic field on the primary CRs is investigated, however the statistics is insufficient to obtain significative conclusions.

The resulting pointing accuracy of the Sun shadow detection is found to be consistent with the expectations.

\begin{acknowledgments}
The authors acknowledge the contribution of an anonymous referee for significantly improving the quality of the analysis. The authors acknowledge the financial support of the funding agencies:
Centre National de la Recherche Scientifique (CNRS), Commissariat \`a
l'\'ener\-gie atomique et aux \'energies alternatives (CEA),
Commission Europ\'eenne (FEDER fund and Marie Curie Program),
Institut Universitaire de France (IUF), LabEx UnivEarthS (ANR-10-LABX-0023 and ANR-18-IDEX-0001),
R\'egion \^Ile-de-France (DIM-ACAV), R\'egion
Alsace (contrat CPER), R\'egion Provence-Alpes-C\^ote d'Azur,
D\'e\-par\-tement du Var and Ville de La
Seyne-sur-Mer, France;
Bundesministerium f\"ur Bildung und Forschung
(BMBF), Germany; 
Istituto Nazionale di Fisica Nucleare (INFN), Italy;
Nederlandse organisatie voor Wetenschappelijk Onderzoek (NWO), the Netherlands;
Council of the President of the Russian Federation for young
scientists and leading scientific schools supporting grants, Russia;
Executive Unit for Financing Higher Education, Research, Development and Innovation (UEFISCDI), Romania;
Ministerio de Ciencia, Innovaci\'{o}n, Investigaci\'{o}n y Universidades (MCIU): Programa Estatal de Generaci\'{o}n de Conocimiento (refs. PGC2018-096663-B-C41, -A-C42, -B-C43, -B-C44) (MCIU/FEDER), Severo Ochoa Centre of Excellence and MultiDark Consolider (MCIU), Junta de Andaluc\'{i}a (ref. SOMM17/6104/UGR and A-FQM-053-UGR18), 
Generalitat Valenciana: Grisol\'{i}a (ref. GRISOLIA/2018/119), Spain; 
Ministry of Higher Education, Scientific Research and Professional Training, Morocco.
We also acknowledge the technical support of Ifremer, AIM and Foselev Marine
for the sea operation and the CC-IN2P3 for the computing facilities.
\end{acknowledgments}


\begin{thebibliography}{99}

\bibitem{Main}
  M. Ageron \textit{et al.}, \href{https://doi.org/10.1016/j.nima.2011.06.103} {Nucl. Instrum. Methods Phys. Res., \\
  Sect. A \textbf{656}, 11 (2011).}
  
\bibitem{cyg}
  D. E. Alexandreas \textit{et al.}, \href{https://doi.org/10.1103/PhysRevD.43.1735} {Phys. Rev. D \textbf{43}, 1735 (1991).}
  
\bibitem{tib}
  M. Amenomori \textit{et al.}, \href{https://doi.org/10.1016/j.astropartphys.2007.05.002} {Astropart. Phys. \textbf{28}, 137 (2007).}
  
\bibitem{bust}
 Yu. M. Andreyev, V. N. Zakidyshev, S. N. Karpov, and V. N. Khodov, \href{https://doi.org/10.1023/A:1021553713199} {Cosmic Res. \textbf{40}, 559 (2002).}
  
\bibitem{casa}
  A. Borione \textit{et al.}, \href{https://doi.org/10.1103/PhysRevD.49.1171} {Phys. Rev. D \textbf{49}, 1171 (1994).}
  
\bibitem{macro}
  M. Ambrosio \textit{et al.}, \href{https://doi.org/10.1016/S0927-6505(03)00169-5} {Astropart. Phys. \textbf{202}, 145 (2003).}
  
\bibitem{soudan}
  J. H. Cobb \textit{et al.}, \href{https://doi.org/10.1103/PhysRevD.61.092002} {Phys. Rev. D \textbf{61}, 092002 (2000).}
  
\bibitem{argo}
  B. Bartoli \textit{et al.}, \href{https://doi.org/10.1103/PhysRevD.85.022002} {Phys. Rev. D \textbf{85}, 022002 (2012).}
  
\bibitem{HAWC}
  A. U. Abeysekara \textit{et al.}, \href{https://doi.org/10.1103/PhysRevD.97.102005} {Phys. Rev. D \textbf{97}, 102005 (2018).}
     
\bibitem{minos} 
  P. Adamson \textit{et al.}, \href{https://doi.org/10.1016/j.astropartphys.2010.10.010} {Astropart. Phys. \textbf{34}, 457 (2011).}
  
\bibitem{ice}
  M. G. Aartsen \textit{et al.}, \href{https://doi.org/10.3847/1538-4357/aaffd1} {Astrophys. J. \textbf{872}, 133 (2019).}
  
\bibitem{Moon_Antares}
   A. Albert \textit{et al.}, \href{https://doi.org/10.1140/epjc/s10052-018-6451-3} {Eur. Phys. J. C \textbf{78}, 1006 (2018).}  
  
\bibitem{reconstructed}
  S. Adrian-Martinez \textit{et al.}, \href{https://doi.org/10.1088/1475-7516/2013/03/006} {J. Cosmol. Astropart. Phys. 006 (2013) 1303.}
  
\bibitem{muon_cand}
 J. A. Aguilar \textit{et al.},\href{https://doi.org/10.1016/j.astropartphys.2011.01.003}  {Astropart. Phys. \textbf{34}, 652 (2011).}
 
\bibitem{energy_threshold}
 Y. Becherini, A. Margiotta, M. Sioli, and M. Spurio, \href{https://doi.org/10.1016/j.astropartphys.2005.10.005} {Astropart. Phys. \textbf{25}, 1 (2006).}
  
\bibitem{mupage}
  G. Carminati, M. Bazzotti, A. Margiotta, and M. Spurio, \href{https://doi.org/10.1016/j.cpc.2008.07.014} {Comput. Phys. Commun. \textbf{179}, 915 (2008).}
 
\bibitem{mupage_description}
H. Yepes-Ramirez \textit{et al.}, \href{https://doi.org/10.1016/j.nima.2012.11.143} {Nucl. Instrum. Methods Phys. Res., Sect. A \textbf{725}, 203 (2013).}
  
\bibitem{RbR}
  L.A. Fusco and A. Margiotta, \href{https://doi.org/10.1051/epjconf/201611602002} {Eur. Phys. J. Web Conf. \textbf{116}, 02002 (2016).}

\bibitem{quality}  
  A. Albert \textit{et al.}, \href{https://doi.org/10.1103/PhysRevD.96.082001} {Phys. Rev. D \textbf{96}, 082001 (2017).}
  
\bibitem{PDG} 
P.A. Zyla \textit{et al.} (Particle Data Group), \href{http://pdg.lbl.gov/2020/reviews/rpp2020-rev-statistics.pdf} {to be published in Prog. Theor. Exp. Phys. \textbf{2020}, 083C01 (2020)}
  
\bibitem{point-source1}
  A. Albert \textit{et al.}, \href{https://doi.org/10.3847/1538-4357/ab7afb} {Astrophys. J. \textbf{892}, 92 (2020).}
  
\bibitem{point-source2}  
  S. Adrian-Martinez \textit{et al.}, \href{http://doi.org/10.1088/2041-8205/786/1/L5} {Astrophys. J. Lett. \textbf{786}, L5 (2014).}

  
\end{thebibliography}
\end{document}